\documentstyle[12pt,epsfig,rotate]{article}
\begin{document}

\setcounter{page}{1}
 \thispagestyle{empty}
%\begin{flushleft}
\begin{center}
{\large\bf
Polarization and spin correlation parameters in proton knockout
reactions from s$_{1/2}$-orbits at 1 GeV }
\end{center}

\begin{center}

 {
O.V.~Miklukho (1), A.Yu.~Kisselev (1), D.A.~Aksenov (1), G.M.~Amalsky (1) ,\\
V.A.~Andreev (1), S.V.~Evstiukhin (1), A.E.~Ezhilov (1), O.Ya.~Fedorov (1),\\
G.E.~Gavrilov (1), D.S.~Ilyin (1), A.A.~Izotov (1), L.M.~Kochenda (1),\\
P.A.~Kravtsov (1),  M.P.~Levchenko (1), D.A.~Maysuzenko (1), V.A.~Murzin (1),\\
T.~Noro (2), D.V.~Novinsky (1), V.A.~Oreshkin (1), A.N.~Prokofiev (1),\\
A.V.~Shvedchikov (1), V.Yu.~Trautman (1), S.I.~Trush (1) and
A.A.~Zhdanov (1)\\}

\end{center}

\begin{center}
%\begin{flushleft}
{\it
(1) St.Petersburg Nuclear Physics Institute, Gatchina, Russia\\
(2) Department of Physics, Kyushu University, Fukuoka, Japan }
%\end{flushleft}
\end{center}
\vspace*{1.cm}

The polarization of the secondary protons (P$_{1,2}$) in the
(p,2p) reaction with the S - shell protons of nuclei $^4$He, $^6$Li,
$^{12}$C, $^{28}$Si, $^{40}$Ca was measured at 1 GeV unpolarized
proton beam. The spin correlation parameters C$_{i,j}$ for the
$^4$He and $^{12}$C targets also were for the first time obtained.
The polarization measurements were performed by means of a
two - arm magnetic spectrometer, each arm of which was equipped
with multiwire - proportional chamber polarimeter.

\vspace*{1cm}

{\bf Comments :} 17 pages, 5 figures, 6 tables.

\vspace*{1cm}

{\bf Category :} Nuclear Experiment (nucl-ex)

\newpage
\normalsize

\begin{center}
{\bf Abstract}
\end{center}

{ The polarization of the secondary protons (P$_{1,2}$) in the
(p,2p) reaction with the S - shell protons of nuclei $^4$He, $^6$Li,
$^{12}$C, $^{28}$Si, $^{40}$Ca was measured at 1 GeV unpolarized
proton beam. The spin correlation parameters C$_{i,j}$ for the
$^4$He and $^{12}$C targets also were for the first time obtained.
The polarization measurements were performed by means of a two -
arm magnetic spectrometer, each arm of which was equipped with
multiwire - proportional chamber polarimeter. This experimental
work was aimed to study a modification of the proton - proton
scattering matrix in the nuclear medium. }
\normalsize
\section{Introduction}
\label{sect1}
 In recent years the question of medium modifications of nucleons and mesons masses
and sizes, and of meson - nucleon coupling constants, and, as
consequence, of a nucleon - nucleon scattering matrix, has received a
great deal of attention [1-6]. These modifications have been
motivated from a variety of theoretical points of view, which include
renormalization effects due to strong relativistic nuclear fields,
deconfinement of quarks, and partial chiral symmetry restoration.

 The  present work is a part of the wide experimental program in the frame of which
 medium - induced modifications of nucleon - nucleon scattering amplitudes are
 studied at PNPI synchrocyclotron with 1 GeV of proton beam energy [7-13].
 A (p,2p) reaction  with nuclei is considered as the proton - proton scattering
 in nuclear matter. In the exclusive (p,2p) experiments the two - arm magnetic spectrometer
 (MAP and NES spectrometers)  is used, the shell structure of
 the nuclei being clearly distinguished. To measure polarization characteristics of the
 reaction, each arm of the spectrometer is equipped with multiwire - proportional chamber
 polarimeter.

 In joint PNPI - RCNP experiments in 2000 - 2002 years, the polarization of both
 secondary protons P$_{1,2}$ in the (p,2p)
 reactions with the 1S - shell protons of nuclei $^6$Li, $^{12}$C and
 with the 2S - shell protons
 of the $^{40}$Ca nucleus was measured at
 the nuclear proton momenta before the (p,2p) interaction
 close to zero [10]. The polarization observed in the experiment
 (as well as the analyzing power A$_y$ in the RCNP (p,2p) experiment
at the 392 MeV polarized proton beam [14,15])
 drastically differs from that calculated in the framework of non - relativistic Plane Wave Impulse
 Approximation (PWIA) and of spin - dependent
  Distorted Wave Impulse Approximation (DWIA)[16], based on free space proton - proton
 interaction.
 This difference was found to be negative and increases monotonously with
 the effective mean nuclear density $\bar\rho$ [14], which is determined by an
 absorption of initial and secondary protons in nucleus matter.
 Note also that the observed small difference between the non - relativistic PWIA and DWIA
 calculations
 pointed out on a small contribution from the trivial depolarization of secondary protons due
 to the final state interaction.
 All these facts strongly indicated a modification of proton - proton
 scattering amplitudes due to a modification of main properties
 of hadrons in nuclear medium. Relativistic calculations have been done to analyze and explain the
 experimental data [10,12].

 The result of the (p,2p) experiment with the $^4$He target, performed in 2004 year [11],
broke the mentioned above monotonous dependence of a difference
between experimental obtained polarization  and that
calculated in the frame of the PWIA on the effective mean nuclear
density $\bar\rho$. This difference for the $^4$He nucleus proved
to be close to that for the $^6$Li nucleus in spite of that the
$^4$He nucleus, according to studies of the elastic
nucleon - nucleus scattering, has the largest mean nuclear density.
On the other hand mentioned above deviation from the PWIA keeps to
be a monotonous function of the S - shell proton binding energy
E$_s$ for all the investigated nuclei. It is possible that in
light nuclei, where nuclear matter is strongly heterogeneous, the
value of $\bar\rho$ does not reflect good enough the scale of
nuclear medium influence on hadron properties and nucleon - nucleon
interaction. The results observed in the (p,2p) experiment
with the light nuclei
at least show that the value of E$_s$ may also be a measure of the
influence of nuclear medium on the pp - interaction.

At present work  polarization of the secondary protons P$_{1,2}$
in the (p,2p) reaction with the 1S - shell protons of the $^{28}$Si
nucleus was measured in the kinematics close to that of the
elastic proton - proton scattering (momenta of nuclear protons
before interaction were close to zero). The goal of the experiment
was to define the relative deviation of experimental polarization
from that calculated in the PWIA is determined by the the
1S - proton binding energy E$_s$ or by the the effective mean
nuclear density $\bar\rho$.
 The mean value of the E$_s$ for the $^{28}$Si
nucleus (50 MeV) is essentially larger then that for the $^{12}$C
nucleus (35 MeV). In the same time the values of the $\bar\rho$
seen in the kinematics of the (p,2p) reactions for these nuclei
are close to each other due to a saturation of the nuclear matter.

This experimental program was extended to measure the spin
correlation parameters C$_{i,j}$  in the (p,2p) reaction with the
1S - shell protons of the $^4$He and $^{12}$C nuclei. The left index
i (i = n, s$^,$) and the right index j (j = n, s$^{,,}$) are
correspond to the forward scattered proton analyzed by the
MAP polarimeter and the recoil proton analyzed by the NES
polarimeter, respectively. Here  {\bf n} is the unit vector
perpendicular to the  scattering plane of the (p,2p) reaction.
Unit vectors {\bf s}$^,$ and {\bf s}$^{,,}$, which lie in the
scattering plane, are concerned to the internal coordinate systems
of the MAP and NES polarimeters.

The main interest was related to measuring the spin correlation
parameter C$_{nn}$  since it's value is the same in the
center - of - mass and laboratory systems, and dos'not distort by
magnetic fields of the two - arm spectrometer due to the proton
anomalous magnetic moment [17]. Since the polarization P and  the
spin correlation parameter C$_{nn}$ depend differently  on the
scattering matrix element [6], measurement of both these
polarization observables in a (p,2p) experiment with nuclei can
give more comprehensive information about modification hadron
properties in nuclear medium.

\section{Experimental method}
The general layout of the experimental setup used to investigate
(p,2p) reaction with nuclei is presented in Fig.~1.

The experiment is performed at the non - symmetric scattering angles
of the final state protons in the coplanar quasi - free scattering
geometry with a complete reconstruction of the reaction
kinematics. The measured secondary proton momenta K$_1$, K$_2$
and scattering angles $\Theta_1$, $\Theta_2$ are used together
with the proton beam energy T$_0$ to calculate nuclear proton
separation energy $\Delta$E and the residual nucleus momentum {\bf
K}$_r$ for each (p,2p) event.  In impulse approximation, the K$_r$
is equal to the momentum (K) of nuclear proton before the
interaction ({\bf K}$_r$=-{\bf K}).

 External proton   beam of the PNPI synchrocyclotron was focused
onto the  target TS of two - arm spectrometer (the magnetic
spectrometers MAP and NES). The beam intensity was monitored by
the scintillation telescope M1, M2, M3 and was about of
5$\cdot$10$^{10}$ protons/(s$\cdot$cm$^2$).

The solid nuclear targets TS made from CH$_2$ (for the setup
calibration), $^6$Li, $^{12}$C, $^{28}$Si, $^{40}$Ca (Table 1) and
the universal cryogenic target with the liquid helium  $^4$He (or
with the liquid hydrogen for calibration) were used in the
experiment [11,18]. Cylindrical aluminium appendix of the
cryogenic target had the following dimensions: diameter - 65 mm,
height - 70 mm, wall thickness - 0.1 mm. The diameter of the beam
spot on the target was less than 15 mm.

The two - arm spectrometer was used for registration of the
secondary protons from the (p,2p) reaction in coincidence and for
measurement of their momenta and outgoing angles. The polarization
of these protons $P_1$ and $P_2$, and the spin correlation
parameters C$_{i,j}$ were measured by the polarimeters located in
the region of focal planes of spectrometers MAP and NES. The
polarimeter of spectrometer MAP (NES) consisted of proportional
chambers PC1$\div$PC4, PC1', PC4' (PC5$\div$PC8, PC5', PC8') and
carbon analyzer A1 (A2).

The main parameters of the two - arm magnetic spectrometer and
polarimeters are listed in Table~2 and Table~3, respectively. The
$\Delta$E resolution of the spectrometer estimated on the elastic
proton - proton scattering with the 22-mm-thick cylindrical CH$_2$
target (see Table~1) was found to be about of 5 MeV (FWHM).

The track information from proportional chambers of both
polarimeters was used in the offline analysis to find the
azimuthal $\phi_1$, $\phi_2$ and polar $\theta_1$, $\theta_2$
angles of proton scattering from the analyzers A1, A2 for each
(p,2p) event. In the case of absence of the accidental coincidence
background (the case of the elastic proton - proton scattering)
the polarization parameters could be found as [19]
\begin{equation}
P_{1,2}=\frac{ 2<\cos\phi_{1,2}> }{ <A(\theta_{1,2},K_{1,2})> }\\,
\end{equation}

\begin{equation}
C_{nn}=\frac{  4<\cos\phi_1\cos\phi_2> }{<A(\theta_1,K_1)><A(\theta_2,K_2)> }\\,
\end{equation}

\begin{equation}
C_{s^,{s^{,,}}}=\frac{  4<\sin\phi_1\sin\phi_2> }{<A(\theta_1,K_1)><A(\theta_2,K_2)> }\\,
\end{equation}

\begin{equation}
C_{n{s^{,,}}}=\frac{  4<\cos\phi_1\sin\phi_2> }{<A(\theta_1,K_1)><A(\theta_2,K_2)> }\\,
\end{equation}

\begin{equation}
C_{s^,{n}}=\frac{  4<\sin\phi_1\cos\phi_2> }{<A(\theta_1,K_1)><A(\theta_2,K_2)> }\\,
\end{equation}

where averaging was made over a set of events within the working
angular range of $\theta_{1,2}$ (see Table~3) for the MAP and NES
polarimeters. $A(\theta_1,K_1)$ and $A(\theta_2,K_2)$, which were
averaged over the same set of events, are the carbon analyzing
power parameterized according to [20] and [21] for the MAP and NES
polarimeter, respectively.

At present work the polarization parameters were estimated by
folding the theoretical functional shape of the azimuthal angular
distribution into experimental one [13], using the CERNLIB MINUIT
package [22] and likelihood $\chi^2$ estimator [23]. This method
permits to realize the control over $\chi^2$ in the case the
experimentally measured azimuthal distribution is distorted due to
the instrumental problems.

The Time-of-Flight (TOF), the time difference between the signals from the
scintillation counters S2 and S4 was measured. This measurement
served to  control  the accidental coincidence background. The
events from four neighboring proton beam bunches were recorded.
Three of them contained the background events only and were used in
the offline analysis to estimate the background polarization
parameters and the background contribution at the main bunch containing
the correlated (p,2p) events.

  The recoil spectrometer NES  was installed at a fixed angle $\Theta_2=53.22^\circ$.
At a given value of the S - shell mean binding energy of nucleus
under investigation, the angular and momentum settings of the MAP
spectrometer and the momentum setting of the NES spectrometer were
chosen to get a kinematics of (p,2p) reaction close to that of the
free elastic proton - proton scattering. In this (p,2p) kinematics,
momentum  {\bf K} of the nuclear proton before the interaction is
close to zero. At this condition the counting rate of the S - shell proton
knockout reaction should be maximal. In Fig.~2 the proton
separation energy spectrum for the (p,2p) reaction on the
$^{28}$Si nucleus, obtained at present work, is presented. As seen
from the figure, even at the preferable condition for the S - shell
proton knockout process, the contribution from the scattering off
the external shell protons is dominant.
\newpage
\begin{figure}
\centering\epsfig{file=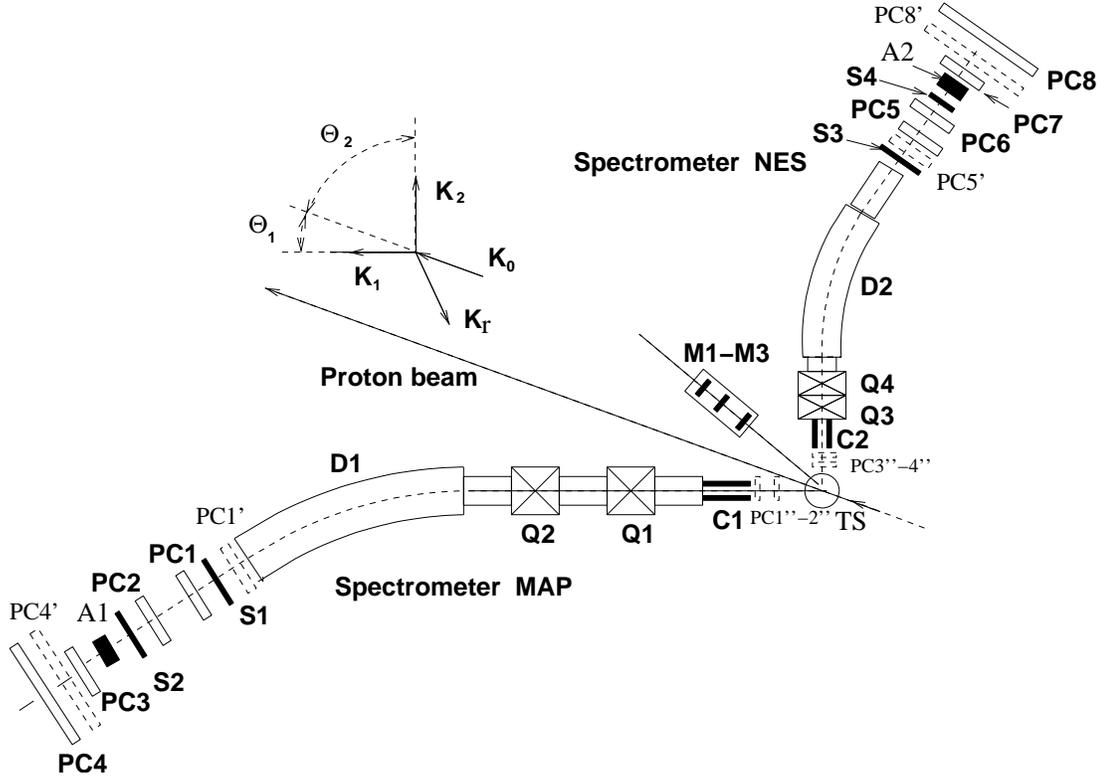, width=1.\textwidth} \caption{The
experimental setup. TS is the target of two - arm spectrometer;
Q1$\div$Q4 are the magnetic quadrupoles; D1, D2 are the dipole
magnets; C1, C2 are the collimators; S1$\div$S4 and M1$\div$M3 are
the scintillation counters; PC1$\div$PC4, PC1', PC4'
(PC5$\div$PC8, PC5', PC8') and A1 (A2) are the proportional
chambers and carbon analyzer of the high - momentum (low -
momentum) polarimeter, respectively; PC1"$\div$PC4" are the
proportional chambers. Shown above is the kinematics for the
(p,2p) reaction.} \label{f_One}
\end{figure}

Measurements of the spin correlation parameters and even of the
polarization in the (p,2p) reaction with heavy nuclei became
possible due the fast proportional chamber readout system
(CROS-3), developed and produced at PNPI [13]. This electronics
allowed to collect the correlation events without distortion at a
high rate of the accidental coincidence background.
{
\begin{table}
\begin{center}
\begin{tabular}{l|c|c}
\multicolumn{3}{c}{\normalsize TABLE 1: Solid target parameters}\\
\multicolumn{3}{l}{ }\\
\hline
\hline
Target               &  Dimensions (mm)           & Isotope concentration ($\%$) \\
\hline
                     &  diameter x high           &\\
CH$_2$               &  22x70                 &         \\
                  & thick x wide x high           & \\
$^6$Li                  & 4.5x12x25             &   99.0 \\
$^{12}$C                & 4.0x18x70            &   98.9 \\
$^{28}$Si               & 6.0x25x70            &   99.9 \\
$^{40}$Ca               & 4.0x10x13            &   97.0 \\
\hline
\hline
\end{tabular}
\end{center}
\end{table}

\begin{table}
\begin{center}
%  \caption{ Parameters of the magnetic spectrometers}
%\label{t_One}
%\vspace*{0.5cm}
\begin{tabular}{l|ll}
\multicolumn{3}{c}{\normalsize TABLE 2: Parameters of the magnetic spectrometers}\\
\multicolumn{3}{l}{ }\\
\hline
\hline
Spectrometer & NES & MAP\\
\hline
Maximum particle momentum & & \\
$(K/Z)^{max}$, GeV/c  & 1.0 & 1.7\\
\hline
Axial trajectory radius $\rho$, m & 3.27 & 5.5\\
\hline
Deflection angle $\beta$, deg. & 37.2 & 24.0\\
\hline
Dispersion in focal plane $D_f$, $mm/\%$ & 24 & 22\\
\hline
Solid angle acceptance $\Omega$, sr & $3.1\cdot 10^{-3}$ &
      $4.0\cdot 10^{-4}$\\
\hline
Momentum acceptance $\Delta K/K$, $\%$ & 8.0 & 8.0\\
\hline
Energy resolution (FWHM), MeV & $\sim 2.0$ & $\sim 1.5$\\
\hline
\hline
\end{tabular}
\end{center}
\end{table}
\begin{table}
\begin{center}
%\caption{Polarimeter parameters}
\begin{tabular}{l|c|c}
\multicolumn{3}{c}{\normalsize TABLE 3: Polarimeter parameters}\\
\multicolumn{3}{l}{ }\\
\hline \hline
Polarimeter                    & NES           & MAP\\
\hline
Carbon block thickness, mm   & 79          & 199 \\
Polar angular range, deg.      & 5$\div$18 & 3$\div$16\\
Average analyzing power        & $\geq$ 0.46   & $\geq$ 0.23\\
\hline \hline
\end{tabular}
\end{center}
\end{table}
}

\newpage
\begin{figure}
\centering\epsfig{file=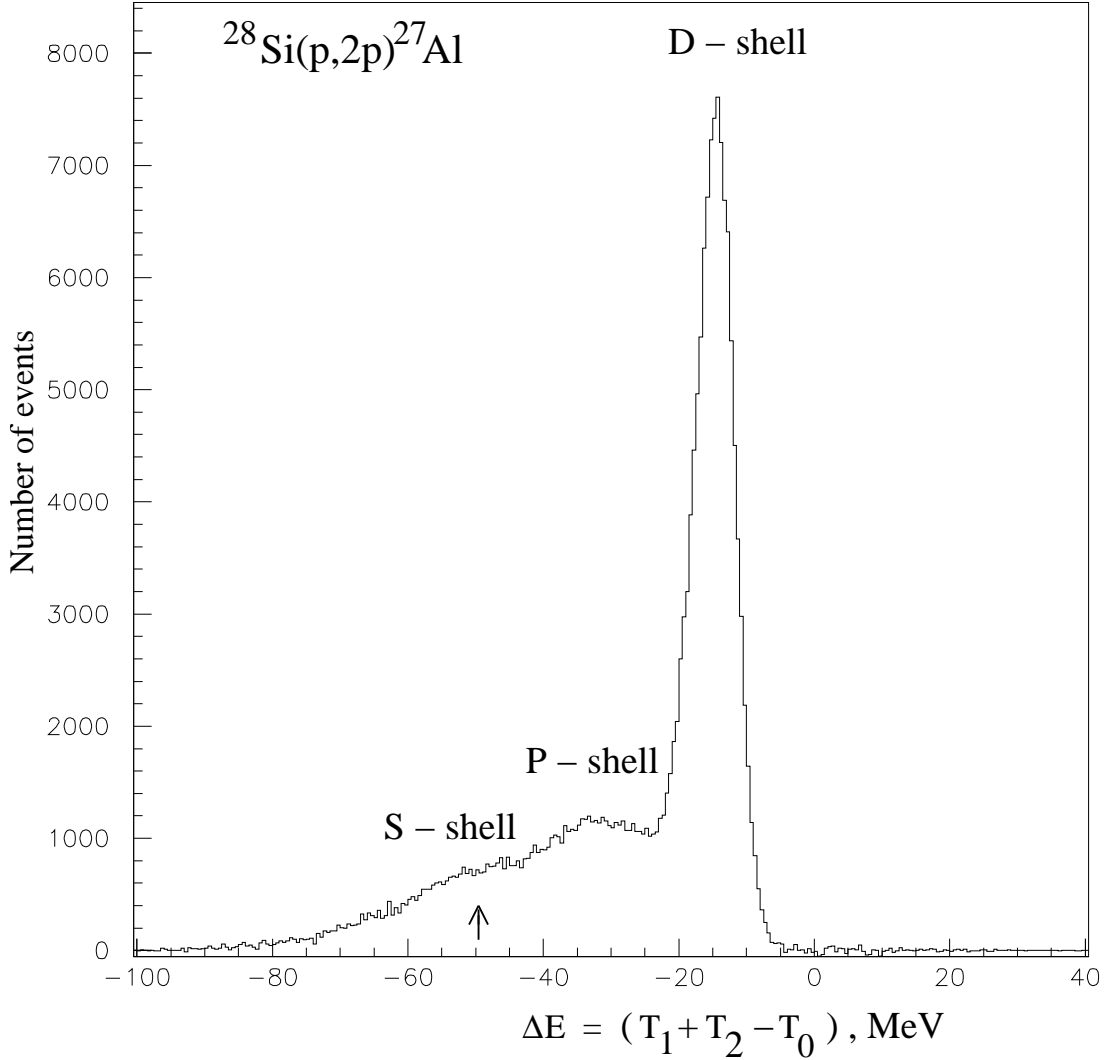, width=1.\textwidth} \caption{
Proton separation energy spectrum for the reaction
$^{28}$Si(p,2p)$^{27}$Al. } \label{f_Two}
\end{figure}

\newpage
\section{Experimental results and discussion}

The measured polarization in the (p,2p) reactions with the S - shell
protons of nuclei $^4$He, $^6$Li, $^{12}$C, $^{28}$Si, $^{40}$Ca
is given in Table~4 (see Appendix). In Fig.~3, the averaged values of the data
with those obtained earlier in [10] are plotted versus of the
S-shell proton binding energy E$_s$ and the effective mean nuclear
density $\bar\rho$, normalized on the saturation nuclear density
$\rho_0\approx$~0.18 fm$^{-3}$ (see also Table~4). The points
($\circ$) and ($\bullet$) in the figure correspond to the
polarization P$_1$ and P$_2$ of the forward scattered protons at
angle $\Theta_1$ = 21$^\circ \div$ 25$^\circ$ (with the kinetic
energy T$_1$ = 745 $\div$ 735 MeV) and the recoil protons
scattered at the angle $\Theta_2\approx$ 53.2$^\circ$ (with the
energy T$_2$ = 205 $\div$ 255 MeV). The points at the E$_s$ = 0
are the polarizations P$_1$ and P$_2$
\begin{figure}
\centering\epsfig{file=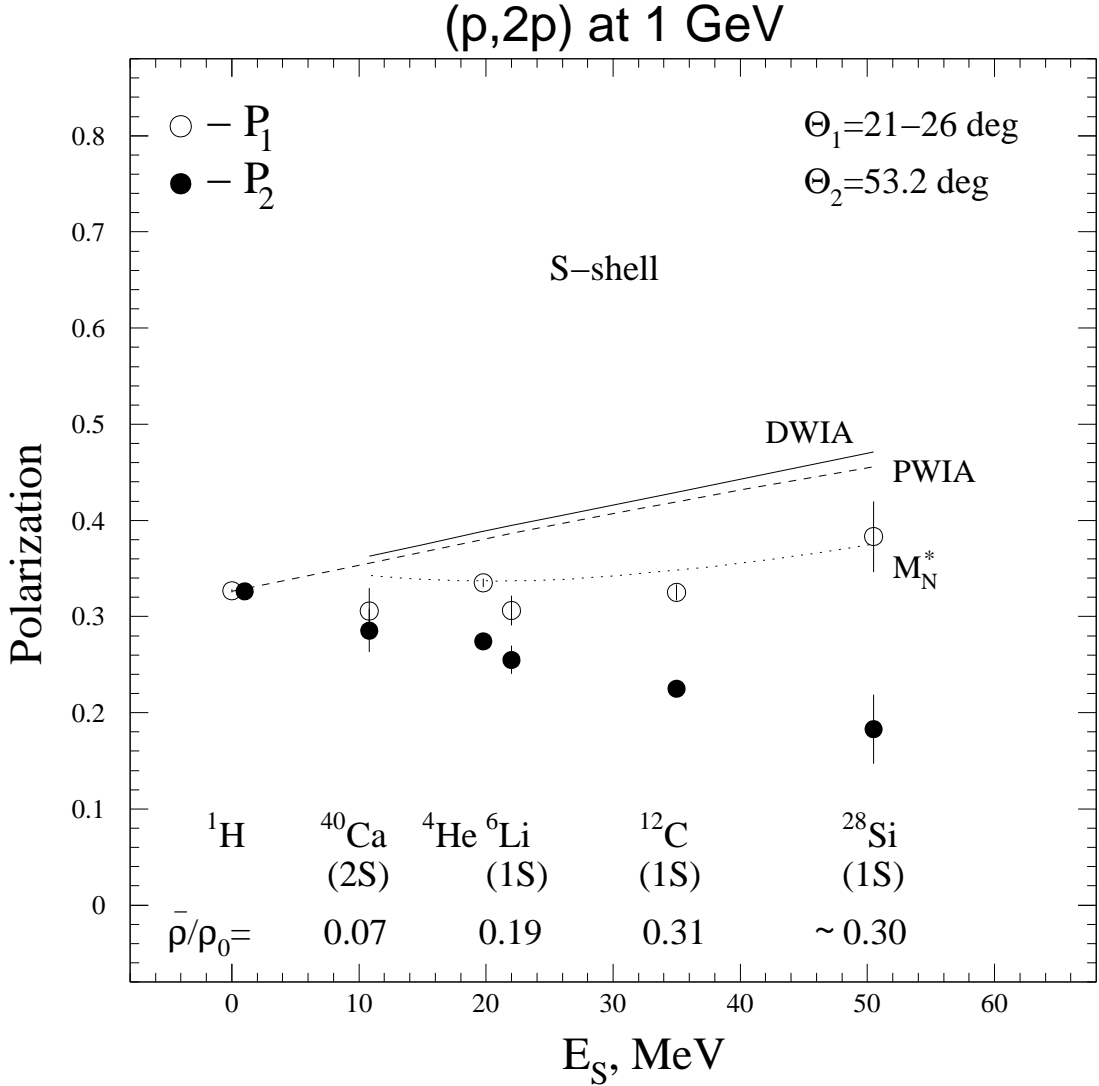,width=1.\textwidth}
\caption{Polarizations P$_1$ and P$_2$  of the  protons scattered
at the angles $\Theta_1$  ($\circ$) and $\Theta_2$ = 53.22$^\circ$
($\bullet$) in the (p,2p) reaction with the S - shell protons of
nuclei at 1 GeV as a function of the mean binding energy E$_s$ and
the effective mean nuclear density, $\bar\rho$ [14], in units of
the saturation density ($\rho_0$ = 0.18 fm$^{-3}$). The points at
E$_s$=0 correspond to the elastic proton - proton scattering
($\Theta_1$ = 26.0$^\circ$). The dashed curve and the solid curve
are the results of calculation in the PWIA  and the DWIA,
respectively, with the NN interaction in free space [16]. The
dotted curve is the DWIA result, in which the relativistic effect
is taken into account in a Schr\"{o}dinger equivalent form [5].}
\label{f_Three}
\end{figure}
in the elastic proton - proton scattering at the angles $\Theta_1$ =
26.0$^\circ$ and $\Theta_1$ = 53.2$^\circ$ ($\Theta_{cm}$ =
62.25$^\circ$). Note that these pp - data were obtained by a
renormalization of the polarimeter analyzing power requiring that
the measured polarization should match the value (P$_{1,2}$ =
0.326) given by the current phase - shift analysis SP07 [25]. The
normalization coefficient was about of 1.06 for both polarimeters.
This correction of the analyzing power was also done  for the
polarization  data obtained in  the (p,2p) experiment with nuclei.

In Fig.~3 the experimental data are compared with the results of
non - relativistic PWIA (plane wave impulse approximation) and DWIA
(distorted wave impulse approximation) calculations employing
 an on - shell factorized approximation.
 The dashed and solid curves, corresponding to PWIA and DWIA, respectively, present
the results of the calculations, which were obtained  using the
computer code THREEDEE [16]. A global optical potential [26],
parametrized in the relativistic framework and converted to the
Shr\"{o}dinger - equivalent form, was used to calculate the
distorted waves of incident and outgoing protons in the case of
DWIA, and a conventional well - depth method was used to construct
bound - state wave function. To calculate free observables in the
density independent NN interaction, the THREEDEE code uses the
1986 Arndt NN phase - shift analysis (SP86) [27]. The results of the
calculations presented in Fig.~3 were normalized on a ratio of the
PWIA predictions obtained with the current phase - shift analysis
SP07 and old one SP86. The value of ratio P(SP07)/P(SP86) was
about of 1.025. Note here, that the $^4$He polarization data were
analyzed only in the framework of the PWIA.

Because the difference between P$_1$ and P$_2$ values in the DWIA
calculations was found to be small, no more than 0.02, only the
P$_1$ values obtained from DWIA are plotted in Fig.~3. As seen
from the figure, the difference between the PWIA and DWIA results
is quite small. This result suggests that the distortion, in a
conventional non - relativistic framework, does not play an
essential role in the polarization for the kinematic conditions
employed in the present work. The final energy prescription [24]
was used for the PWIA and DWIA calculation. We also found that the
difference between the initial and final prescriptions was small
in these kinematic region. The strong positive slope of the
polarizations predicted by these calculations (see Fig.~3) is caused
by the kinematic effects of the binding energy of the struck
proton.

The differences between the polarizations P$_1$, P$_2$  calculated
in the PWIA  and  those  measured in the (p,2p) reaction with nuclei
$^{40}$Ca, $^6$Li, $^{12}$C are monotonically increasing functions
of the effective mean density (see Fig.~3) [14]. The relative
polarization difference (Pexp - Pia)/Pia is shown in Fig.~4. This
difference for the $^{28}$Si nucleus, for the forward proton
polarization P$_1$ at least, confirms also that the depolarization
effect is determined by the effective mean nuclear density
$\bar\rho$. Indeed, the values of the relative differences for
$^{12}$C and $^{28}$Si nuclei ($\circ$ - points) are practically
equal to each other, these nuclei having the same value of the
$\bar\rho$ and strongly different the S - shell mean binding
energy E$_s$. This observation provides further evidence that
there exists a nuclear medium effect.
\begin{figure}
\centering\epsfig{file=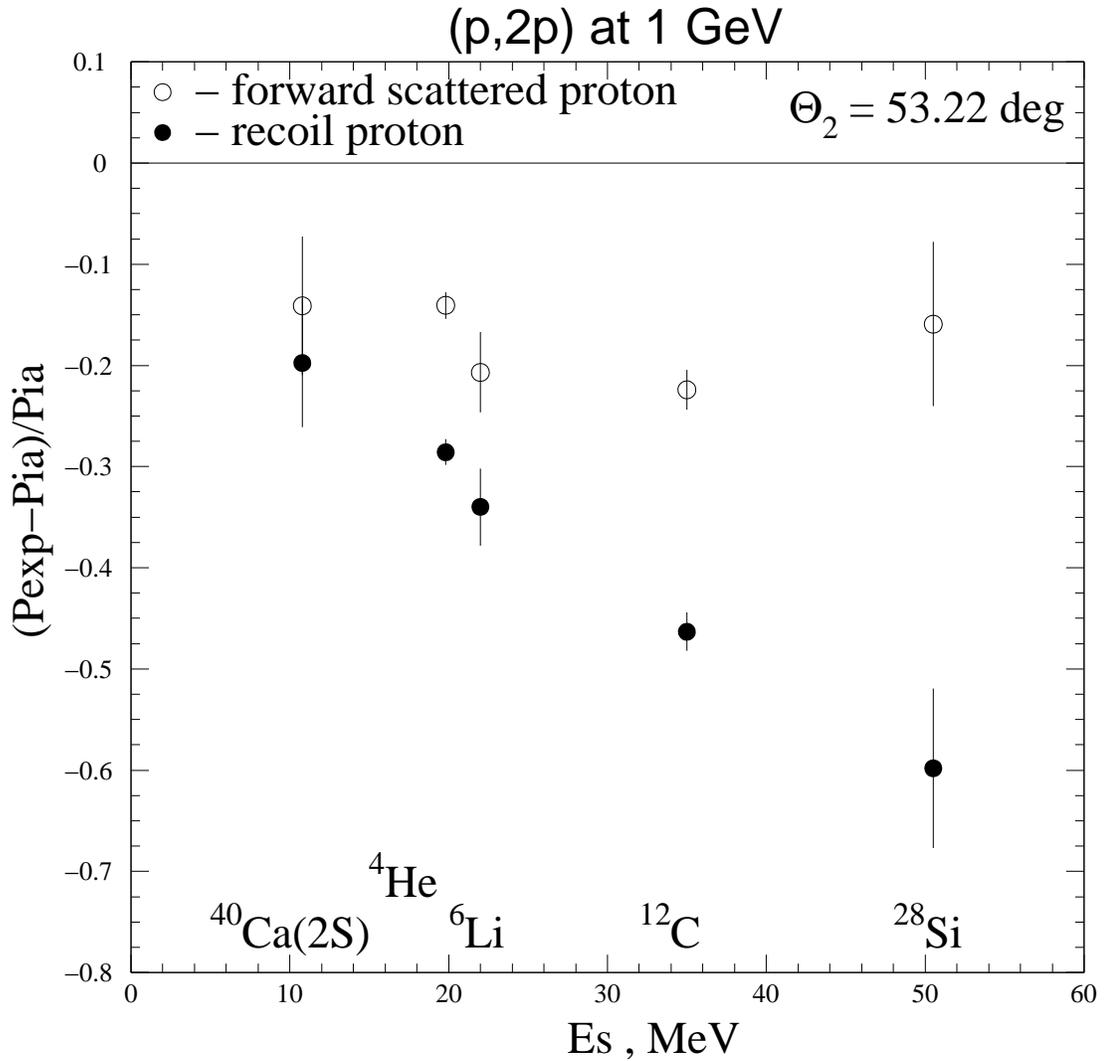,width=1.\textwidth} \caption{ A
relative deviation of the polarization  observed in the (p,2p)
reaction with S - shell protons of nuclei from that calculated in
the PWIA. The points ($\circ$) and ($\bullet$) correspond to
 the forward scattered proton at the angle
$\Theta_1$ = 21.0$^\circ$$\div$ 25.08$^\circ$ and the recoil
proton scattered at the angle $\Theta_2$ = 53.22$^\circ$,
respectively (see Fig.~3).}
\label{f_Four}
\end{figure}
As seen from~Fig.~3-4, there is a systematic difference between
the P$_1$ and P$_2$ values, though they have the same values in
the case of elastic pp - scattering. Possible origins of the
difference between these values include non - relativistic and
relativistic distortions (though the former is excluded if the
present DWIA calculations are valid), contributions of multi -
step processes, and even nontrivial modification of nucleons in
the nuclear field.

In Fig.~3 the experimental data are compared with a theoretical
result for the case when a relativistic effect, the distortion of
the nucleon spinor, is taken into account. The calculation was
carried out in the Shr\"{o}edinger equivalent form [10] using the
THREEDEE code [16]. More specifically, this calculation consists
of a non - relativistic DWIA calculation with a nucleon - nucleon
t - matrix, that is modified in the nuclear potential following a
procedure similar to that proposed by Horowits and Iqbal [5]. In
this approach a modified NN interaction in nuclear medium is
assumed due to the effective nucleon mass (smaller than the free
mass) which affects the Dirac spinors used in the calculations of
the NN scattering matrix. A linear dependence of the effective
mass of nucleons on the nuclear density was assumed in the
calculations. As seen from the Fig.~3, this relativistic approach
gives the results (the dotted curve) close to the experimental values of
the forward  scattered proton polarization P$_1$ in the (p,2p)
reactions with nuclei at the transfered momenta q = 3.2$\div$3.7 fm$^{-1}$
(see Table~4).

Another possible medium effect is the modifications of exchanged
meson masses and meson - nucleon coupling constants in the NN
interaction. Krein et al. have shown in the relativistic Love -
Franey model (RLF) that these modifications cause significant
changes in the spin observables which include suppression of $A_y$
[6]. A such type of modification was investigated in [12] using
our experimental data on polarization in the (p,2p) reaction with
the S - shell protons of the $^{12}$C nucleus obtained in a wide
range of the momentum transfer q [10]. Note that at present work
we essentially improved a statistic accuracy of the polarization
measurements in the (p,2p) reaction with the $^{12}$C nucleus at
the q = 3.4 fm$^{-1}$.

In the present work, the spin correlation parameters C$_{i,j}$ in
the (p,2p) reactions with the S - shell protons of the $^4$He and
$^{12}$C nuclei were for the first time measured using an
unpolarized 1 GeV proton beam.

Since the polarization P and the spin correlation parameter
C$_{nn}$ depend differently  on the scattering matrix elements
[6], measurement of both these polarization observables in a
(p,2p) experiment with nuclei can give more comprehensive
information about modification of the hadron properties in nuclear
medium.

The results of the C$_{i,j}$ measurement in the (p,2p) reaction
with nuclei are given in Fig.~5 (Table~5). In Table~5 the measured
mean values of the C$_{i,j}$ for the accidental coincidence
background, obtained in investigating the (p,2p) reaction with
nuclei $^{12}$C and $^{28}$Si, are also presented.
\begin{figure}
\centering\epsfig{file=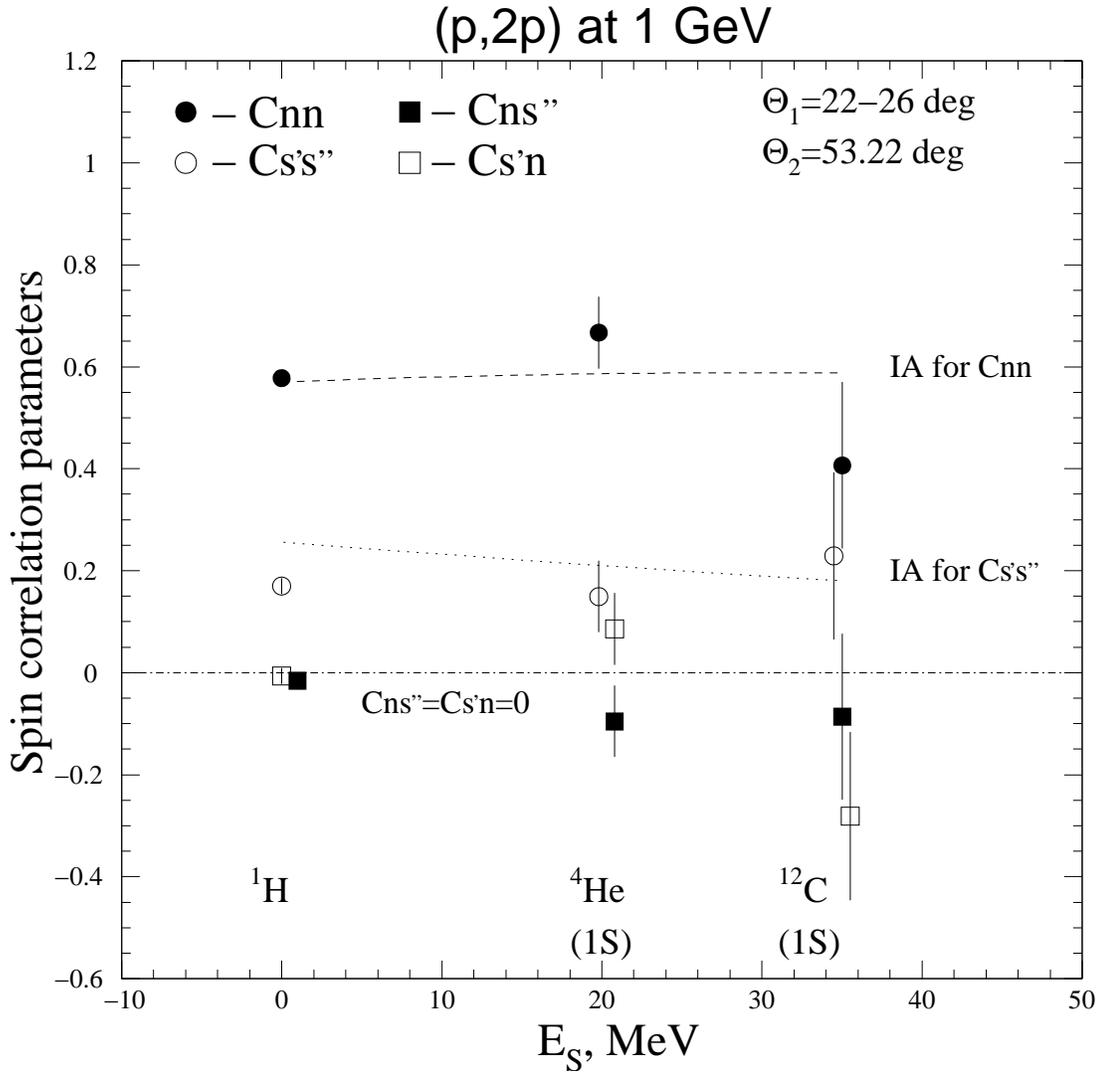,width=1.\textwidth} \caption{ Spin
correlation parameters C$_{i,j}$ in the (p,2p) reaction at 1 GeV
with the S - shell protons of the $^4$He and $^{12}$C nuclei at
the secondary proton scattering angles $\Theta_2$ = 53.2$^\circ$,
$\Theta_1$ = 24.2$^\circ$ and $\Theta_2$ = 53.2$^\circ$,
$\Theta_1$ = 22.7$^\circ$, respectively. The points at E$_s$ = 0
correspond to the free elastic proton-proton scattering
($\Theta_1$ = 26.0$^\circ$, $\Theta_{cm}$ = 62.25$^\circ$),
obtained in 2009 year experiment (Table~6). The dashed curve and
the dotted curve are the results of the PWIA calculation of the
C$_{nn}$ and C$_{s^,{s^{,,}}}$ spin correlation parameters.}
\label{f_Five}
\end{figure}
In Fig.~5 the dashed and dotted curves are correspond to the PWIA
calculations for the C$_{nn}$ and C$_{s^,{s^{,,}}}$ spin
correlation parameters. In these calculations the current Arndt
phase - shift analysis (SP07) was used [25]. The  C$_{s^,{s^{,,}}}$
parameter was found by taking into account it's distortion in the
magnetic fields of the MAP and NES spectrometers due to an
anomalous proton magnetic moment [17]. The points at the mean
binding energy value of E$_s$ = 0 correspond to the free
elastic proton - proton scattering (see Table~6).

As seen from Fig.~5, the differences between the C$_{nn}$ values
measured in the (p,2p) experiment with the nuclei  and those
calculated in the PWIA are within the statistical error bars.

The measured value of the C$_{s^,{s^{,,}}}$ parameter in the elastic proton - proton scattering
strongly differs from the SP07 prediction. This can be related to a lack of the spin correlation
parameter data from the elastic pp - scattering experiments
at 1 GeV in the base of the current phase - shift analysis.

Due to the parity conservation in the free elastic proton - proton
scattering the the spin correlation parameters C$_{ns{^{,,}}}$ and
C$_{s^,{n}}$ should be equal to zero. In Fig.~5, this  is
confirmed by the experimental data at the E$_s$=0. For  a (p,2p)
reaction with nuclei the parity in the pp - interaction system can
be violated since there
exists a residual nucleus in the knockout
process. However in this case, according to the Pauli principle, a
relation C$_{ns{^{,,}}}$ = -C$_{s^,{n}}$ for the pp - interaction
system should be performed [28].

\section{Acknowledgments}

This work is partly supported
by the Grant of President of the Russian Federation for Scientific School,
Grant-3383.2010.2.

The authors are grateful to PNPI 1~GeV proton accelerator staff
for stable beam operation. We thank members of PNPI HEP
Radio-electronics Laboratory for providing the CROS-3 proportional
chamber readout system.

Also, the authors would like to express their gratitude to A.A.~Vorobyov and S.L.~Belostotski
 for their support and fruitful discussions.

\newpage

\section{Appendix:}

\begin{center}
\small
\begin{tabular}{c|c|c|c|c|c|c|c|c}
\multicolumn{9}{p{\textwidth}}{\normalsize TABLE 4:  Polarization
of secondary protons $P_1$ and $P_2$ produced in the (p,2p)
reaction at 1 GeV with the S - shell protons of nucleus at lab.
angle $\Theta_1$ and $\Theta_2$}\\
\multicolumn{9}{l}{ }\\
\hline
\hline
Nucleus & $\Theta_1$ & $\Theta_2$ & $T_1$ & $T_2$ & $ q $ & $P_1$ & $P_2$  &
$\bar\rho/\rho_0$\\
     &deg. & deg. & MeV & MeV & fm$^{-1}$ &       &       &               \\
\hline
$^4$He (1S) & 24.21 & 53.22 & 738 & 242 &     & 0.336$\pm$0.005 & 0.274$\pm$0.005 &
\\
$<$$^4$He (1S)$>$ & 24.21 & 53.22 & 738 & 242 & 3.6 & 0.335$\pm$0.005 & 0.274$\pm$0.004 &
\\
\hline
$^6$Li (1S) & 24.0 & 53.22 & 738 & 241 &     & 0.309$\pm$0.026 & 0.247$\pm$0.023 &
\\
$<$$^6$Li (1S)$>$ & 24.0 & 53.25 & 739 & 239 & 3.6 & 0.306$\pm$0.015 & 0.255$\pm$0.015 & 0.19
\\
\hline
$^{12}$C (1S) & 22.71 & 53.22 & 746 & 219 &     & 0.329$\pm$0.009 & 0.227$\pm$0.008 &
\\
$<$$^{12}$C (1S)$>$ & 22.71 & 53.22 & 746 & 219 & 3.4 & 0.325$\pm$0.008 & 0.225$\pm$0.008 & 0.31
\\
\hline
$^{28}$Si (1S) & 21.0 & 53.22 & 746 & 204 & 3.2 & 0.383$\pm$0.037 & 0.183$\pm$0.036 & $\sim$0.30
\\
\hline
$^{40}$Ca (2S) & 25.05 & 53.22 & 733 & 256 &     & 0.306$\pm$0.037 & 0.304$\pm$0.033 &
\\
$<$$^{40}$Ca (2S)$>$ & 25.08 & 53.15 & 734 & 255 & 3.7 & 0.306$\pm$0.024 & 0.286$\pm$0.023 & 0.07
\\
\hline
\hline
\end{tabular}
\vspace{0.5cm}
\normalsize
\end{center}

\begin{center}
\small
\begin{tabular}{c|c|c|c|c|c}
\multicolumn{6}{p{\textwidth}}{\normalsize TABLE 5: Spin
correlation parameters C$_{ij}$ in the (p,2p) reaction at 1 GeV
with the 1S - shell protons of the $^4$He and $^{12}$C nuclei at
lab.~angles $\Theta_1$  and $\Theta_2$ = 53.22$^\circ$. The line
"Background" corresponds to the measured mean values of the C$_{ij}$
for the accidental coincidence background, obtained in
investigating
the (p,2p) reaction with nuclei $^{12}$C and $^{28}$Si}\\
\multicolumn{6}{l}{ }\\
\hline
\hline
Nucleus & $\Theta_1$ & C$_{nn}$   & C$_{s^,}$$_{s^{,,}}$ & C$_{ns^{,,}}$ & C$_{s^,}$$_n$
\\
        & deg.       &          &          &          &
\\
\hline
$^4$He  & 24.21 & 0.667$\pm$0.070 & 0.150$\pm$0.070 & -0.095$\pm$0.070 & 0.086$\pm$0.070
\\
\hline
$^{12}$C & 22.71 & 0.407$\pm$0.163 & 0.229$\pm$0.164 & -0.086$\pm$0.163 & -0.281$\pm$0.165
\\
\hline
Background&      & -0.003$\pm$0.020 & 0.005$\pm$0.020  & 0.019$\pm$0.020  & 0.004$\pm$0.020
\\
\hline
\hline
\end{tabular}
\vspace{0.5cm}
\normalsize
\end{center}

\begin{center}
\small
\begin{tabular}{c|c|c|c|c|c|c}
\multicolumn{7}{p{\textwidth}}{\normalsize TABLE 6: Spin
correlation parameters C$_{ij}$ in the elastic proton - proton
scattering at 1 GeV at lab. angles $\Theta_1$ = 26.0$^\circ$   and
$\Theta_2$ = 53.22$^\circ$ ($\Theta_{cm}$ = 62.25$^\circ$),
obtained in 2007-2010 years. The current phase - shift analysis
predicts the C$_{nn}$ value of 0.57. The statistical errors in the
measurements of the polarizations P$_1$ and
P$_2$ are also given}\\
\multicolumn{7}{l}{ }\\
\hline
\hline
Year &  C$_{nn}$  & C$_{s^,}$$_{s^{,,}} $ & C$_n$$_{s^{,,}}$ & C$_{s^,}$$_n$ & $\delta$P$_1$ & $\delta$P$_2$
\\
\hline
2007 & 0.587$\pm$0.021 & 0.115$\pm$0.021 &  0.005$\pm$0.021 & 0.080$\pm$0.021 & 0.0015  & 0.0012
\\
\hline
2008 & 0.584$\pm$0.014 & 0.195$\pm$0.014 & -0.004$\pm$0.014 & 0.008$\pm$0.014 & 0.0010 & 0.0009
\\
\hline
2009 & 0.577$\pm$0.016 & 0.170$\pm$0.016 & -0.016$\pm$0.016 & -0.006$\pm$0.016 & 0.0015  & 0.0013
\\
\hline
2010 & 0.455$\pm$0.052 & 0.162$\pm$0.052 & -0.050$\pm$0.052 &  0.009$\pm$0.052 & 0.0041  & 0.0036
\\
\hline
\hline
\end{tabular}
\vspace{0.5cm}
\normalsize
\end{center}

\end{document}